\documentclass[aps,pre,twocolumn,showkeys]{revtex4}

\usepackage{amsmath,graphicx,bm,amssymb,mathrsfs}

\newcommand{\Ueff}{U_{\mathrm{eff}}}
\def \SL {S_{\rm L}}
\def \SN {S_{\rm N}}
\def \CL  {C_{\rm L}}

\def \FN  {F_{\rm N}}
\def \FL  {F_{\rm L}}

\def \xx {{\bm x}}

\def \del2z {\partial^{2}_{z}}

\def \uu {{\bm u}}

\def \JJ {{\bm J}}

\def \curl {{\bm \nabla} \times}
\def \dive {{\bm \nabla}\cdot}

\newcommand{\Eq}[1]{Equation~(\ref{#1})}

\newcommand{\bra}[1]{\left\langle #1\right\rangle}

{}

\def \BB {\bm B}

\def \curl {{\bm \nabla}\times}

\def \BB {\bm B}

 \def \xS {\xi_{\rm S}}
 \def \xP {\xi_{\rm P}}
 \def \xc {\xi_{\rm crit}}
 \def \rk {r_{\kappa}}

\def\drawing #1 #2 #3 {
\begin{center}
\setlength{\unitlength}{1mm}
\begin{picture}(#1,#2)(0,0)
\put(0,0){\framebox(#1,#2){#3}}
\end{picture}
\end{center} }

\begin{document}

NORDITA 2018-086

\medskip

\title{Kazantsev dynamo in turbulent compressible flows}

\author{Marco Martins Afonso$^{1}$, Dhrubaditya Mitra$^2$ and Dario Vincenzi$^{3}$}

\affiliation{$^{1}$Centro de Matem\'atica (Faculdade de Ci\^encias) da Universidade do Porto, Rua do Campo Alegre 687, Porto 4169-007, Portugal\\
 $^{2}$Nordita, KTH Royal Institute of Technology and Stockholm University, Sweden\\
 $^{3}$Universit\'e C\^ote d'Azur, CNRS, LJAD, Nice 06100, France}

\title{Kazantsev dynamo in turbulent compressible flows}


\begin{abstract}
We consider the kinematic fluctuation dynamo problem in a flow that is random, white-in-time, with both solenoidal and potential components. This model is a generalization of the well-studied Kazantsev model. If both the solenoidal and potential parts have the same scaling exponent, then, as the compressibility of the flow increases, the growth rate decreases but remains positive. If the scaling exponents for the solenoidal and potential parts differ, in particular if they correspond to typical Kolmogorov and Burgers values, we again find that an increase in compressibility slows down the growth rate but does not turn it off. The slow down is, however, weaker and the critical magnetic Reynolds number is lower than when both the solenoidal and potential components display the Kolmogorov scaling. Intriguingly, we find that there exist cases, when the potential part is smoother than the solenoidal part, for which an increase in compressibility increases the growth rate. We also find that the critical value of the scaling exponent above which a dynamo is seen is unity irrespective of the compressibility. Finally, we realize that the dimension $d=3$ is special, since for all other values of $d$ the critical exponent is higher and depends on the compressibility.
\end{abstract}

\maketitle

\section{Introduction} \label{sect:introduction}

Astrophysical objects typically have magnetic fields over many ranges
of scales. The growth and saturation of these magnetic fields are the
subject of dynamo theory. Most astrophysical flows are turbulent, hence
the astrophysically relevant magnetic fields are generated by
turbulent dynamos. Broadly speaking, we understand two different kinds of
dynamo mechanism in turbulent fluids \cite{BS05}: (a) one that generates magnetic
fields whose characteristic length scales are significantly larger than
the energy-containing length scales of the flow --- large-scale dynamos;
and (b) one that generates small-scale magnetic fields --- the
fluctuation dynamo. The small-scale tangled magnetic fields in the
interstellar medium (ISM) or the Sun are supposed to be generated by the
fluctuation dynamo. If we are interested in only the initial growth
of the magnetic field but not its saturation, then we can ignore the
Lorentz force by which the magnetic field acts back on the flow and
reduce a non-linear problem to a linear one --- the kinematic dynamo
problem. A pioneering work on the kinematic fluctuation dynamo was by
Kazantsev \cite{K68}, who approximated the velocity field by a
random function of space and time. In particular, the velocity field
is assumed to be statistically stationary, homogeneous and isotropic
with zero divergence (under the latter assumption the
velocity is called solenoidal or incompressible).
It is also assumed to be short-correlated in
time, more specifically white-in-time, and its correlation function in
space is assumed to have a power-law behaviour with an exponent
$0\leqslant \xi\leqslant 2$.
By virtue of the white-in-time nature of the velocity field, it is
possible to write down a closed equation for the equal-time two-point
correlation function of the magnetic field. Consequently, it is
possible to show that a dynamo exists --- the magnetic energy grows
exponentially in time --- iff the exponent $\xi > 1$~\cite{V96}.
This result is an example of an anti-dynamo theorem.
The Kazantsev model has played an important role in our understanding of
the kinematic fluctuation dynamo excited by turbulent flows {\it \`a la}
Kolmogorov, see e.g.\ the review by Brandenburg and Subramanian
\cite{BS05} and references
therein. The same model for the velocity field has been extensively
used to study the scaling behaviour and intermittency of advected
passive scalar fields --- 
the Kraichnan model~\cite{Kraichnan} of passive-scalar turbulence,
see e.g.\ \cite{FGV01} for a review.

In this paper we are interested in a generalization of the Kazantsev
model to {\it compressible} turbulence, which is relevant for the fluctuation
dynamo in the ISM \cite{F11,F14,F16}. To the best of our knowledge, the earliest attempt to generalize the
Kazantsev model to compressible flows was by Kazantsev {\em et al.}~\cite{KRS85}, 
who modelled the flow not by a white-in-time process but by 
random weakly-damped sound waves, with energy concentrated on
a single wave number, to find that the growth rate of the dynamo is
proportional to the fourth power of the Mach number. 
More recently, Rogachevskii and Kleeorin \cite{RK97} generalized the Kazantsev model to
compressible flows, preserving its white-in-time nature, but adding a
potential (irrotational) component to the velocity field. Both the solenoidal and
the potential components are assumed to have the same scaling
exponent, $0\leqslant \xi\leqslant 2$.  In this model, Rogachevskii and Kleeorin
\cite{RK97} found that the inclusion of 
compressible modes always makes it more difficult to excite the dynamo.
Later Schekochihin and Kulsrud \cite{SK01} and Schekochihin {\em et al.} \cite{SBK02} 
studied the same problem for the case of
smooth velocity fields, $\xi=2$, which corresponds to the 
limit of large magnetic Prandtl numbers.
With the purpose of understanding the magnetic-field growth in ISM,
Schober {\em et al.} \cite{Schober12} incorporated compressibility
into a generalization of the Kazantsev model previously 
proposed by Subramanian \cite{S97}.
However, a tensor function can be the correlation 
of an isotropic vector field iff
the corresponding longitudinal and normal three-dimensional spectra
are non-negative \cite[p.~41]{MY75}:
the velocity correlation prescribed by Subramanian \cite{S97}
and then by Schober {\em et al.}
\cite{Schober12} does not satisfy such condition.
In addition, in the model of Schober {\em et al.}~\cite{Schober12}
the scaling exponent of the velocity correlation and the degree of 
compressibility are not independent parameters.
In particular, for the Kolmogorov scaling the velocity is incompressible; this limit of the Kazantsev
model has been studied in many papers before. The focus of our study is in contrast
the dependence of the dynamo effect on the degree of compressibility.

Recent direct numerical simulations of compressible 
fluid turbulence~\cite{F10,F13,Wangetal13,Wangetal18} have
shown that the spectra of the flow can be separated into a
solenoidal and a potential part, where the solenoidal part scales
with an exponent quite close to the classical Kolmogorov result for
incompressible turbulence --- the energy spectrum of the solenoidal
part scales with exponent $-5/3$ in the inertial range ---
and the scaling exponent for the potential part is close to that 
of the turbulent Burgers equation --- the energy spectrum of the
potential part scales with an exponent of $-2$ in the
inertial range.  How will the Kazantsev dynamo problem change if we
use two independent scaling exponents for the solenoidal and the
potential parts of the energy spectrum of the velocity field? This
is the main task we set for ourselves in this paper.  

We end this introduction by a warning to our reader: 
because of the popularity of the Kazantsev model as an analytically
tractable model for the turbulent dynamo,
the literature on it
is large and diverse. We have so far cited only
those papers that are directly relevant to our work. For a detailed
introduction we suggest several reviews \cite{ZRS90,FGV01,BS05,TCB12} and references therein.

\section{Model} \label{sect:model}

We model the plasma as a conducting fluid at scales where
the magnetohydrodynamics (MHD)
equations are valid. Consequently the magnetic field, $\BB(\xx,t)$, evolves
according to the induction equation~\cite{Cha61}
\begin{equation}
\partial_t \BB = \curl(\uu\times\BB - \kappa \JJ)  \/,
\label{eq:MHD}
\end{equation}
where $\JJ = \curl\BB$ is the current, $\kappa$ the magnetic
diffusivity, and $\uu(\xx,t)$ the velocity. Furthermore, Maxwell's equations 
imply $\dive\BB = 0$. 
Two dimensionless numbers characterize the MHD equations:
the Reynolds number $\mathit{Re}\equiv VL/\nu$ and the magnetic Reynolds number
$\mathit{Re}_{\rm m}\equiv VL/\kappa$, where $V$ is a typical large-scale velocity,
$L$ is the correlation length of the velocity field, and $\nu$ the
kinematic viscosity of the fluid.
The ratio of $\mathit{Re}_{\rm m}$ and $\mathit{Re}$ is the magnetic Prandtl 
number $\mathit{Pr}\equiv\nu/\kappa$.

Instead of solving the
momentum equation for the velocity, as is usual in MHD, the Kazantsev
model assumes that the velocity
$\uu$ is a statistically stationary,
homogeneous, isotropic and parity invariant, Gaussian random
field. 
The statistics of a Gaussian field is entirely defined by its mean
and second-order correlation. In the Kazantsev model, the mean
of $\uu$ is zero and the correlation is
\begin{equation}
\label{eq:velocity}
\langle u_i(\bm x+\bm r,t)u_j(\bm x,t')\rangle = 
\mathscr{D}_{ij}(\bm r)\delta(t-t'),
\qquad i,j=1,\dots,d,
\end{equation}
where $d$ is the spatial dimension of the flow.
In this paper, we use $d=3$ 
except in 
Section~\ref{sect:general-d}, where
results for general $d$ are presented.
The velocity field is also assumed to be Galilean invariant, hence it
is useful to introduce the second-order structure function 
$S_{ij}(\bm r)=\mathscr{D}_{ij}(0)-\mathscr{D}_{ij}(\bm r)$,
which is a Galilean invariant quantity that describes the statistics of the velocity increments.
In view of statistical isotropy and parity invariance, $S_{ij}(\bm r)$
takes the form\footnote{Note that our definitions of the structure functions
differ from those of Monin and Yaglom \cite{MY75} by a factor of~2.} \cite{R40,MY75}
\begin{equation}
\label{eq:structure-function}
S_{ij}(\bm r)= \SN(r)\delta_{ij}+[\SL(r)-\SN(r)]\hat{r}_i \hat{r}_j,
\end{equation}
where $\hat{r}_i=r_i/r$ and $\SN(r)$ and $\SL(r)$ are 
called the normal (also known as transverse or lateral) and
longitudinal second-order structure functions,
respectively.

\subsection{Solenoidal random flows} \label{sec:solran}

Kazantsev \cite{K68}
considered the three-dimensional sole\-noidal case ($\nabla\cdot\bm u=0$)
and the limits of vanishing $\mathit{Pr}$
and infinite $\mathit{Re}$ and $\mathit{Re}_{\rm m}$,
in which the velocity structure functions are scale invariant:
\begin{equation}
\SL(r)=2Dr^{\xi},\quad \SN(r)=(\xi+2)Dr^{\xi}.
\label{eq:kazan}
\end{equation}
The positive constant $D$ determines the magnitude
of the fluctuations of the velocity increments, while
the scaling exponent $\xi$ varies between 0 and 2, the latter value
describing a spatially smooth velocity field.
For the above choice of the structure functions, the spectrum of
the velocity $\bm u$ takes the form $E(k)\propto k^{-1-\xi}$.

We remind the reader that,
owing to the $\delta$-correlation in time,
some care should be taken in comparing the velocity field $\bm u$, defined in
\eqref{eq:velocity}, with a turbulent flow.
Indeed, the eddy diffusivity of a three-dimensional turbulent flow
scales as $r^{4/3}$, which yields Richardson's law
for the 
separation $R(t)$ between two fluid particles:
$\langle R^{2}(t)\rangle\sim t^3$ \cite{F95}.
In the $\delta$-correlated field $\bm u$, the eddy diffusivity scales
as $r^{\xi}$ and fluid particles separate according to the
following law: $\langle R^{2}(t)\rangle\sim t^{2/(2-\xi)}$ \cite{FGV01}.
Therefore, 
in order for the $\delta$-correlated velocity field to reproduce
Richardson's law, 
the scaling exponent $\xi$ must be taken equal to $4/3$.
Analogously,
if the structure functions of a time-correlated three-dimensional
incompressible flow 
scale as $r^{\alpha}$ (or equivalently its spectrum
scales as $k^{-1-\alpha}$)
the eddy diffusivity of such a flow 
behaves as $r^{1+\alpha/2}$ and the associated law of fluid-particle
dispersion is $\langle R^{2}(t)\rangle\sim t^{4/(2-\alpha)}$.
In this case, the time-correlated flow and the $\delta$-correlated 
one yield the same scaling for the eddy diffusivity, and hence the same
law of Lagrangian dispersion, if $\xi=1+\alpha/2$
\cite[pp.~926--927]{FGV01}.

\subsection{Compressible random flows}

One way to include the effect of compressibility into the Kazantsev
model is to modify, in $d$-dimensions,  
$\SL(r)$ and $\SN(r)$ as follows~\cite{EKR95,CLM99,GV00}:
\begin{equation}
\label{eq:structure}
\SL(r)=Dr^{\xi}(d-1)(\wp\xi+1),\ \SN(r)=Dr^{\xi}(-\wp\xi+\xi+d-1),
\end{equation}
by introducing a new parameter $\wp$ which is the degree of compressibility of the velocity field
and varies between 0 (solenoidal, or incompressible, flow) and 1 (potential, or
irrotational, flow).
In this model both the solenoidal and the potential parts of the flow
have the same scaling exponent.

Two regimes can be identified depending on the value of $\wp$
\cite{GV00}. In the regime of weak compressibility ($\wp< d/\xi^2$),
fluid particles still separate according to the law
$\langle R^{2}(t)\rangle\sim t^{2/(2-\xi)}$. Hence the same argument
as for the incompressible case can be repeated to connect 
$\xi$ and the scaling exponent $\alpha$ of the structure functions of a
time-correlated flow, which yields $\xi=1+\alpha/2$. 
For instance, if a three-dimensional time-correlated flow
displays the Burgers scaling ($\alpha=1$), then 
the same law of Lagrangian dispersion is obtained
in the $\delta$-correlated velocity field
if $\xi=3/2$.

In the regime of strong compressibility ($\wp\geqslant d/\xi^2$), by contrast,
the low powers of the particle separation decay in time,
while the high ones grow but more slowly
than in the weakly compressible case. The reason for this behaviour is
the collapse of Lagrangian trajectories due to compressibility effects.
Therefore, the dimensional argument used above to connect
$\xi$ and $\alpha$ does not hold in the regime of strong compressiblity.
Moreover, this latter regime can only
be observed for $d\leqslant 4$, since $\wp\leqslant 1$.

\section{Results}

\subsection{Closed equation for the correlation function of the magnetic field}

Given the velocity field the task is to calculate the equations obeyed
by the correlation functions of the magnetic field, where the averages
are calculated over the statistics of the velocity field. 
The magnetic field is assumed to have the same spatial statistical symmetries as
the velocity field. Its correlations are thus written as
\begin{equation}
\label{eq:magnetic-correlation}
C_{ij}(\bm r,t)= \CL(r,t)\delta_{ij}+\dfrac{r}{d-1}\partial_r \CL(r,t)
(\delta_{ij}-\hat{r}_i\hat{r}_j),
\end{equation}
where a single scalar function $\CL(r,t)$
is sufficient to define $C_{ij}(\bm r,t)$
thanks to the solenoidality of the magnetic field \cite{R40,MY75}. 

By virtue of the Gaussian and white-in-time nature of the velocity
field, it is possible to write a closed equation for the evolution of
$C_{ij}(\bm r,t)$ in a
straightforward manner, by using the induction equation \eqref{eq:MHD}
and then averaging over the statistics of the velocity field. 
To obtain a closed equation, however, we need to calculate averages of 
a triple product: that of the velocity field, of the magnetic field and of
its spatial derivative. This has been obtained by many different methods. 
A well established approach employs
a result known as ``Gaussian integration by parts'':
for a Gaussian, not-necessarily white-in-time, vector-valued 
noise $\bm z(t)$, with components $z_j(t)$, and its arbitrary
functional $F(\bm z)$, 
\begin{equation}
\bra{F(\bm z)z_j(t)} = \int
\mathrm{d}s\bra{z_j(t)z_k(s)}
     \bra{\frac{\delta F}{\delta z_k(s)}}\/,
\end{equation}
see e.g. Section~4.2 in \cite{ZJ96} for a proof.
At the next step we need to integrate
\eqref{eq:MHD} formally to obtain
the magnetic field as a function of the velocity field and then to
calculate the necessary functional derivatives. Then we have to take
the limit $s\to t$. As the noise correlation becomes singular in this
limit, we need to replace the Dirac delta function by a regularised
even function and then take the limits. This regularisation is equivalent
to using the Stratonovich prescription for the noise.
This method has been used extensively in dynamo theory
(see, e.g., \cite{V96,SK01,SBK02,MB12,SA16}) to calculate similar or higher-order
correlation functions. Hence we skip the details of the derivation and
directly write down the result 
(see also \cite{SBK02} and references therein):
%\begin{multline}
%\label{eq:trace}
%\partial_t\CL=(2\kappa+\SL)\partial^2_r\CL+\left[\frac{2(d+1)}{r}\kappa
%+\partial_r\SL+\frac{2(d-1)}{r}\SL+\frac{3-d}{r}\SN\right]\partial_r\CL\\
%+\frac{d-1}{r}\left[\partial_r\SL+\partial_r\SN+\frac{d-2}{r}(\SL-\SN)\right]\CL.
%\end{multline}
\begin{equation}
\label{eq:trace}
\begin{split}
\partial_t\CL&=(2\kappa+\SL)\partial^2_r\CL+\left[\frac{2(d+1)}{r}\kappa
+\partial_r\SL\right.
\\
&\qquad\left.+\frac{2(d-1)}{r}\SL+\frac{3-d}{r}\SN\right]\partial_r\CL
\\
&\quad+\frac{d-1}{r}\left[\partial_r\SL+\partial_r\SN+\frac{d-2}{r}(\SL-\SN)\right]\CL.
\end{split}
\end{equation}

\subsection{Schr\"odinger formulation} \label{sect:schroedinger}

One of Kazantsev's main results is the formulation of the turbulent
dynamo effect as the trapping of a quantum particle in a one-dimensional potential
whose shape is determined by the functional form of the velocity structure functions \cite{K68}.
This formulation leads to an appealing interpretation of the dynamo effect.

\Eq{eq:trace} is a partial differential equation with one space and
one time dimension. By the substitution
\begin{equation}
\label{eq:substitution}
\begin{split}
 \CL(r,t)&=\psi(r,t)r^{-(d-1)}[2\kappa+\SL(r)]^{-1/2}\times
 \\
 &\qquad\times\exp\left[\frac{d-3}{2}\int r^{-1}\frac{2\kappa+\SN(r)}{2\kappa+\SL(r)}\mathrm{d}r\right],
\end{split}
\end{equation}
\eqref{eq:trace} is turned into an imaginary-time
Schr\"odinger equation with space-dependent mass
for the function $\psi(r,t)$ \cite{SBK02}:
\begin{equation}
\label{eq:psi-time}
m(r)\partial_t \psi=\partial^2_r\psi-m(r)U(r)\psi,
\end{equation}
with
\begin{equation}
m(r)=\frac{1}{2\kappa+\SL(r)}
\label{eq:mr}
\end{equation}
and
%\begin{multline}
%\label{eq:potential}
%U(r)=\frac{\partial^2_r\SL(r)}{2}-\frac{[\partial_r\SL(r)]^2}{4[2\kappa+\SL(r)]}+\frac{(d-3)^2[2\kappa+\SN(r)]^2}{4r^2[2\kappa+\SL(r)]}\\
%+\frac{(3d-5)[2\kappa+\SN(r)-r\partial_r\SN(r)]}{2r^2}.
%\end{multline}
\begin{equation}
\label{eq:potential}
\begin{split}
U(r)&=\frac{\partial^2_r\SL(r)}{2}\!-\!\frac{[\partial_r\SL(r)]^2}{4[2\kappa+\SL(r)]}\!+\!\frac{(d-3)^2[2\kappa+\SN(r)]^2}{4r^2[2\kappa+\SL(r)]}
\\
&\quad+\frac{(3d-5)[2\kappa+\SN(r)-r\partial_r\SN(r)]}{2r^2}.
\end{split}
\end{equation}
The function $\psi(r,t)$ has the same time dependence as $\CL(r,t)$.
We therefore
seek solutions of the form $\psi(r,t)=\psi_\gamma(r)
e^{-\gamma t}$, where $\psi_\gamma(r)$ are the eigenstates of the 
Schr\"odinger operator in \eqref{eq:psi-time}
and $\gamma$ the associate energies. 
The magnetic correlation grows in time if there exist negative-energy
($\gamma<0$) eigenstates.
If so, the energy of the ground state, $\gamma_0$,
yields the asymptotic growth rate $\vert\gamma_0\vert$.%
\footnote{Note that here the magnetic growth is studied 
at nonzero $\kappa$. 
Since the $\kappa\to 0$ limit is singular \cite{CFKV99}, 
the asymptotic growth rate
in a perfectly conducting fluid ($\kappa=0$) cannot be deduced
from the results described here by letting $\kappa$ tend to zero.
}
From \eqref{eq:psi-time}, $\gamma$ can be written as:
\begin{equation}
\gamma=\dfrac{\int m(r)U(r)\psi^2_\gamma(r)\mathrm{d}r +\int [\psi'_\gamma(r)]^2\mathrm{d}r}%
{\int m(r)\psi_\gamma^2(r)\mathrm{d}r}.
\end{equation}
Since $m(r)>0$ for all $r$ \cite{SBK02},
$\gamma$ is negative iff the effective potential
$U_{\mathrm{eff}}(r)\equiv m(r)U(r)$ admits negative energies.
Thus the problem of the growth of the magnetic correlations is mapped into
that of the existence of negative-energy states for
the one-dimensional potential $\Ueff(r)$ --- the energy for the ground
state of the potential $\Ueff(r)$ is equal to the growth rate of the
fastest growing mode in the kinematic dynamo problem.

\subsection{Solenoidal random flows}

In the original Kazantsev model, the flow is assumed to be 
solenoidal and the longitudinal and normal structure functions
have the same scaling exponent as given in \eqref{eq:kazan}.
The effective potential is obtained by 
substituting $d=3$ and the expressions of $\SL$ and $\SN$
from \eqref{eq:kazan} in \eqref{eq:trace}, 
and simplifying (e.g. \cite{V96}):
%\begin{align}
%\label{eq:potential-incompressible}
%\begin{split}
%\Ueff(r) & =\frac{8\kappa^2-2(\xi^2+3\xi-8)\kappa
%  Dr^{\xi}-(3\xi^2+6\xi-8)D^2r^{2\xi}}{4r^2(\kappa+Dr^{\xi})^2}\\
%&=\frac{8\rk^{2\xi}-2(\xi^2+3\xi-8)\rk^{\xi}r^{\xi}-(3\xi^2+6\xi-8)r^{2\xi}}{4r^2(\rk^{\xi}+r^{\xi})^2},
%\end{split}
%\end{align}
\begin{equation}
\label{eq:potential-incompressible}
\begin{split}
\Ueff(r)&=\frac{1}{4r^2(\kappa+Dr^{\xi})^2}\left[8\kappa^2-2(\xi^2+3\xi-8)\kappa Dr^{\xi}\right.
\\
&\qquad\left.-(3\xi^2+6\xi-8)D^2r^{2\xi}\right]\\
&=\frac{8\rk^{2\xi}-2(\xi^2+3\xi-8)\rk^{\xi}r^{\xi}-(3\xi^2+6\xi-8)r^{2\xi}}{4r^2(\rk^{\xi}+r^{\xi})^2},
\end{split}
\end{equation}
where in the second step we have introduced the diffusive scale,
$r_\kappa\equiv(\kappa/D)^{1/\xi}$, at which
the diffusive and the advective terms in the induction equation~\eqref{eq:MHD}
balance each other. 
For $r \ll \rk$, $\Ueff(r)\sim 2/r^2$ and is therefore repulsive. 
For $r\gg r_\kappa$, $\Ueff(r)\sim -(3\xi^2/4+3\xi/2-2)/r^2$; hence
the effective potential is negative at large $r$.
By examining the shape of $\Ueff(r)$, it is possible to conclude that the
critical value of $\xi$ for the kinematic dynamo effect is $\xc=1$
\cite{K68} (see also \cite{V96}). For $\xi<1$, $\Ueff(r)$ is indeed
everywhere greater than a potential of the form $-c/r^2$ with $c<1/4$,
which does not have any negative-energy eigenstates \cite{LL58}.
Hence the same holds for $\Ueff(r)$, and consequently no dynamo exists.
For $\xi>1$, the effective potential behaves
at large distances as $\Ueff(r)\sim -c/r^2$ with $c>1/4$.
A potential with such a large-$r$ behaviour has a discrete spectrum 
containing an infinite number of negative-energy levels \cite{LL58},
so the magnetic correlation can grow exponentially.

For a smooth flow ($\xi=2$), 
the value of the growth rate is known analytically:
$\vert\gamma_0\vert=15D/2$ \cite{K68,GCS96,CFKV99},
and the magnetic correlation decays as $\CL(r)\sim r^{-5/2}$
\cite{CFKV99}.
For a spatially rough flow, $1<\xi<2$, 
the analytical calculation of $\vert\gamma_0\vert$ is a difficult task.
It was obtained numerically in \cite{K68} and \cite{V02},
and an analytical estimate was given by Arponen and Horvai \cite{AH07}. The growth rate
increases as a function of $\xi$,
i.e. a greater degree of
spatial regularity of the velocity field favours the magnetic growth.
In addition,
the magnetic correlation decays as a stretched exponential of the 
spatial separation with stretching exponent $1-\xi/2$ \cite{V02}.

The incompressible case was generalized to $d$ dimensions 
by Gruzinov {\em et al.} \cite{GCS96} and Vergassola \cite{V96}. In particular, for $d=2$, $\Ueff(r)$
is repulsive everywhere and hence there cannot be dynamo effect, in accordance
with Zel'dovich's anti-dynamo theorem \cite{Z57}.

\subsection{Compressible random flows} \label{sect:3D}

Next we consider the model of compressible flow given by 
\eqref{eq:structure}, where both the solenoidal and the potential 
components of the structure functions scale with the same exponent. 
Upon substitution of \eqref{eq:structure} into \eqref{eq:mr} and~\eqref{eq:potential},
the mass and potential functions are found to 
have the form
\begin{equation}
m(r)=\frac{1}{2\kappa+Dr^{\xi}(d-1)(\wp\xi+1)}
\end{equation}
and
\begin{equation}
\label{eq:potential-d-wp}
U(r)=\dfrac{a_0+a_1 r^\xi+a_2r^{2\xi}}%
{4r^2[2\kappa+Dr^\xi(d-1)(\wp\xi+1)]},
\end{equation}
where the coefficients $a_0$, $a_1$, $a_2$ are given in appendix~\ref{app:b}.

\begin{figure}
\centering
\includegraphics[width=\columnwidth]{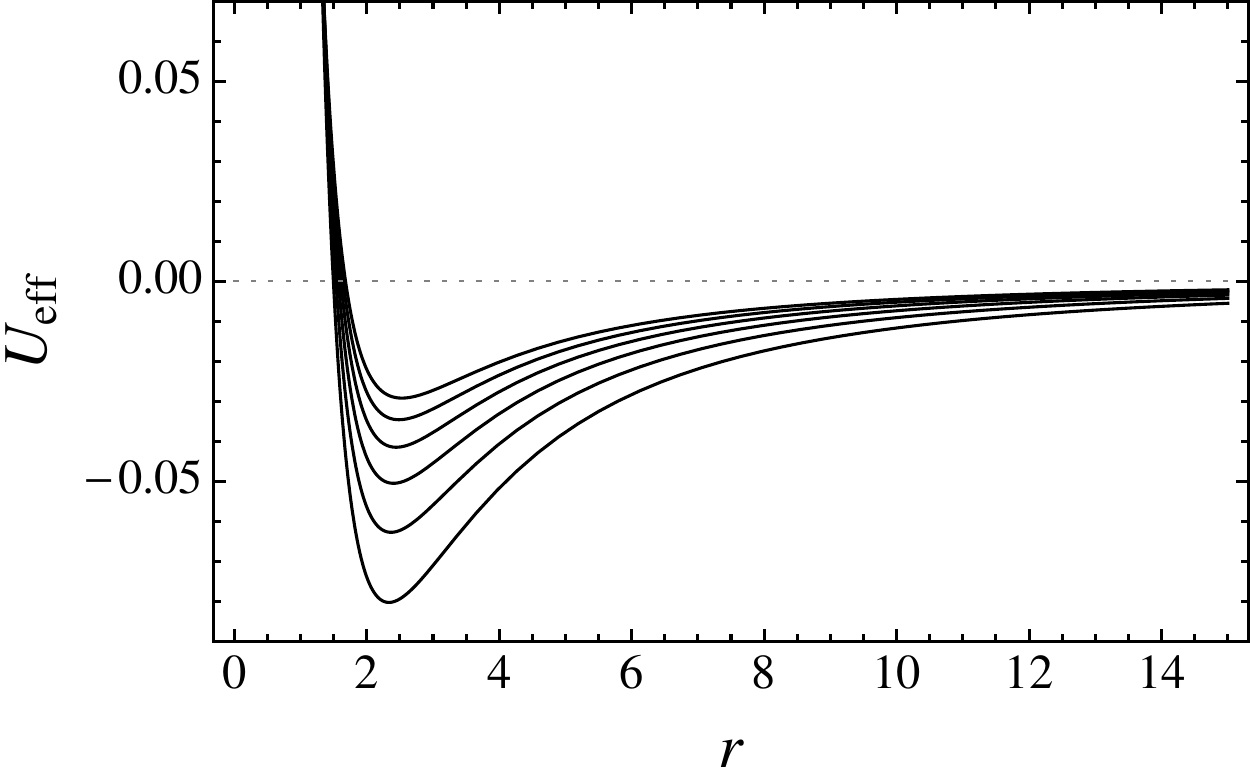}
\caption{Effective potential for $\xi=4/3$ and, from bottom to top, $\wp=0,0.2,0.4,0.6,0.8,1$.}
\label{fig:poteff}
\end{figure}

Let us examine the $d=3$ case.
At small distances ($r\ll r_\kappa$), 
the asymptotic behaviour of $\Ueff(r)$ is unchanged compared to the
incompressible case.
For $r\gg r_\kappa$,
\begin{equation}
\Ueff(r)\sim\frac{\wp\xi^3+2\wp\xi^2-3\xi^2-4\wp\xi-6\xi+8}{4(\wp\xi+1)r^2}\;.
\end{equation}
By studying the form of $\Ueff(r)$ as in Section~\ref{sect:schroedinger},
it can be shown that the critical value of $\xi$ for the dynamo
effect is not affected by the degree of compressibility, i.e. $\xc=1$ for all values
of $\wp$ \cite{RK97}. Indeed, for $\xi<1$, $\Ueff(r)$ behaves
asymptotically as $-c/r^2$ with $c<1/4$, and both $a_0$ and $a_1$ are positive
for all values of $\wp$. Hence $\Ueff(r)>-c/r^2$ with $c<1/4$ for all $r$,
and it does not admit negative-energy eigenstates.
For $\xi>1$, $\Ueff(r)\sim -c/r^2$ with $c>1/4$, and an infinite number
of states with negative energy can exist.
Figure~\ref{fig:poteff} shows the effect of compressibility on the shape of the effective potential.

\begin{figure}
\includegraphics[width=\columnwidth]{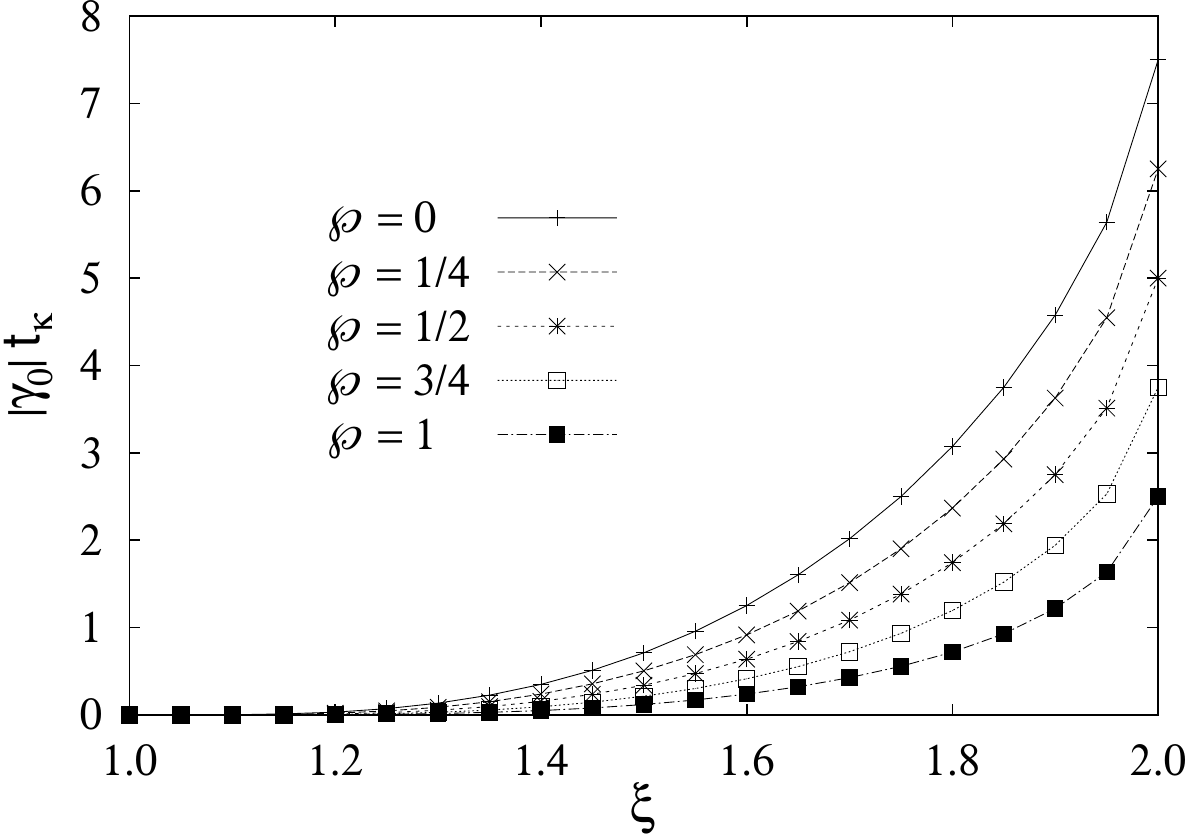}
\\[3ex]
\includegraphics[width=\columnwidth]{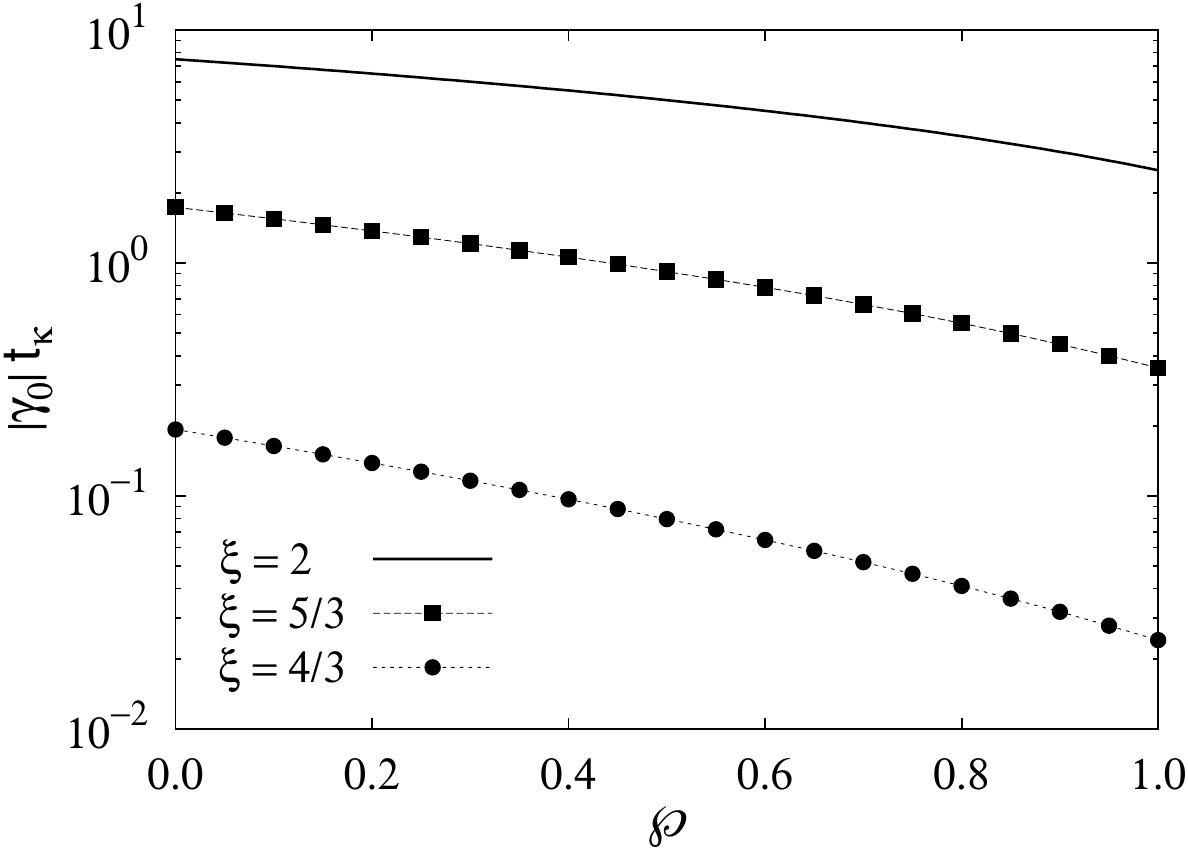}
%\includegraphics[width=0.5\columnwidth]{plot-xi.pdf}%
%\hfill\includegraphics[width=0.5\columnwidth]{plot-wp.pdf}
\caption{Magnetic growth rate multiplied by the diffusive time 
as a function of:
%(left)
(top)
the scaling exponent $\xi$ of the
velocity structure functions, and
%(right)
(bottom)
the degree of compressibility $\wp$ of the
flow. The solid line in the
%right
bottom
panel shows the analytical expression for
$\xi=2$~\cite{SBK02}.}
\label{fig:rate}
\end{figure}

Even though $\xc$ does not vary with $\wp$, 
for $\xi>1$ the magnetic growth is affected by the degree of compressibility of the flow.
For $\xi=2$, the growth rate is
$\vert\gamma_0\vert=(15/2-5\wp)D$, while
at large separations and long times
the correlation $\CL(r,t)$ behaves, up to logarithmic corrections, as
$r^{-5/2}\exp(\vert\gamma_0\vert t)$
\cite{SBK02}.

For $1<\xi<2$, to our knowledge
no analytical expression for the growth rate $\vert\gamma_0\vert$ 
is available.
We calculate it numerically by applying to
\eqref{eq:psi-time} a variation--iteration
method that is similar to techniques used to find the largest eigenvalue
of very large matrices. This method is
described in detail in appendix \ref{app:a}. 
The result is shown in Figure~\ref{fig:rate},
where we plot $\vert\gamma_0\vert t_\kappa$ versus $\xi$ 
for several values of $\wp$;
here $t_\kappa\equiv \rk^2/\kappa$ is the time scale
associated with magnetic diffusion.
We find that, irrespective of $\wp$,
the growth rate is positive for all values of $\xi>1$
and increases with $\xi$, i.e., as the spatial regularity of the 
flow improves. 
Note that, for $\wp=1$, the values of $\xi$ such that
$\xi>\sqrt{3}$ are in the strongly compressible regime (see 
Sect.~\ref{sect:model}).
For a fixed $\xi$, the effect of compressibility is
always to decrease the growth rate, i.e., to make the dynamo weaker. 
Rogachevskii and Kleeorin \cite{RK97} tried to estimate the growth rate by asymptotic matching;
they also found the same qualitative effect of compressibility. 

We find out the behaviour of $\CL(r,t)$ for large $r$ by
replacing $m(r)$ and $U(r)$ in \eqref{eq:psi-time}
with their asymptotic expressions. The resulting equation for
$\psi_\gamma(r)$ is a Bessel differential equation, whose non-diverging
solution is
\begin{equation}
\begin{split}
\psi_\gamma(r)&\propto\sqrt{r}\,K_b[\beta\sqrt{\vert\gamma\vert/D}\,r^{1-\xi/2}]
\\
&=\sqrt{r}\,K_b\left[\beta\sqrt{\vert\gamma\vert t_\kappa}\left(\frac{r}{\rk}\right)^{1-\xi/2}\right],
\end{split}
\end{equation} 
where $\beta=b\sqrt{2/(\wp\xi+1)}$ and
$K_b$ is the modified Bessel function of the second kind
of order $b=1/(2-\xi)$.
This solution behaves as 
$\exp[-\beta\sqrt{\vert\gamma\vert t_\kappa}(r/r_\kappa)^{1-\xi/2}]$ at large $r$.
Furthermore, the long-time behaviour
of the magnetic correlation is determined by the ground state 
$\psi_{\gamma_0}(r)$. From \eqref{eq:substitution} 
we therefore conclude that,
for $r\gg r_\kappa$ and $t\gg t_\kappa$, 
\begin{equation}
\CL(r,t) \propto r^{-2-\xi/2}
\exp\left[{-\beta\sqrt{\vert\gamma_0\vert t_\kappa}\left(\frac{r}{\rk}\right)^{1-\xi/2}}\right]
e^{\vert\gamma_0\vert t}.
\end{equation}
As far as the spatial dependence of the magnetic correlation is
concerned,
we thus recover the 
stretched-exponential behaviour of the incompressible case; the 
degree of compressibility of the flow only affects $\vert\gamma_0\vert$, and
hence the rate at which the stretched exponential function decays in space, but not
the stretching exponent.

\subsection{Solenoidal--potential decomposition} \label{sect:two-exponents}

The results of the previous section indicate that the spatial
regularity of the flow favours the magnetic growth, whereas
compressibility hinders it. 
It has now been realized that for turbulence at moderate Mach numbers
the solenoidal and potential parts of the flow
have different scaling exponents; more precisely,
the solenoidal part of the velocity
spectrum displays the Kolmogorov scaling
$k^{-5/3}$, while the potential one displays the Burgers scaling $k^{-2}$
\cite{F10,F13,Wangetal13,Wangetal18}.
It is therefore interesting to 
investigate the interplay between spatial regularity and compressibility
in the case where the potential part of the flow is more regular
than the solenoidal one. 
We thus consider the following velocity structure functions (see
\cite{MY75}, p.~57,
for the solenoidal--potential decomposition of an isotropic random field):
\begin{flalign}
\label{eq:sl-two-exponents}
\SL(r)&=2(1-\wp)D_{\rm S}r^{\xi_{\rm S}}+2 \wp D_{\rm P}(1+\xi_{\rm P})r^{\xi_{\rm P}},
\\
\label{eq:sn-two-exponents}
\SN(r)&=(1-\wp)D_{\rm S}
(2+\xi_{\rm S})r^{\xi_{\rm S}}+2\wp D_{\rm P}r^{\xi_{\rm P}},
\end{flalign}
where $\xS$ and $\xP$ are the scaling exponents of the 
solenoidal ($\wp=0$) and 
potential ($\wp=1$) components, respectively, and 
$D_{\rm P}$ and $D_{\rm S}$ are positive coefficients.
Varying the degree of compressibility thus interpolates linearly
between a solenoidal flow with scaling exponent $\xS$ and
a potential flow with exponent $\xP$.

The mass and the potential functions
can be calculated by inserting 
$\SL(r)$ and $\SN(r)$ from \eqref{eq:sl-two-exponents}
and \eqref{eq:sn-two-exponents} into \eqref{eq:mr}
and~\eqref{eq:potential}:
\begin{align}
\begin{split}
m(r)&=\frac{1}{2[\kappa+(1-\wp)D_{\mathrm{S}}r^{\xi_{\mathrm{S}}}+\wp D_{\mathrm{P}}r^{\xi_{\mathrm{P}}}]},
\end{split}&
\\[2mm]
\begin{split}
U(r)&=-\frac{[(1-\wp)\xi_{\mathrm{S}}D_{\mathrm{S}}r^{\xi_{\mathrm{S}}-1}+\wp\xi_{\mathrm{P}}(1+\xi_{\mathrm{P}})D_{\mathrm{P}}r^{\xi_{\mathrm{P}}-1}]^2}{2[\kappa+(1-\wp)D_{\mathrm{S}}r^{\xi_{\mathrm{S}}}+\wp(1+\xi_{\mathrm{P}})D_{\mathrm{P}}r^{\xi_{\mathrm{P}}}]}
\\
&\quad+\frac{4\kappa}{r^2}+(1-\wp)(1-\xi_{\mathrm{S}})(4+\xi_{\mathrm{S}})D_{\mathrm{S}}r^{\xi_{\mathrm{S}}-2}
\\
&\quad+\wp(4-5\xi_{\mathrm{P}}+\xi_{\mathrm{P}}^3)D_{\mathrm{P}}r^{\xi_{\mathrm{P}}-2}.
\end{split}&
\end{align}
The product of $m(r)$ and $U(r)$ yields the effective potential.
In the diffusive range, $r\ll\rk$, $\Ueff(r)\sim 2/r^2$
independently of the values of $\xS$ and $\xP$; 
the dynamics is indeed dominated by magnetic diffusion in this range.
If $\xS > \xP$ ($\xS < \xP$) the asymptotic behaviour of $\Ueff(r)$ is the same as
that of a solenoidal flow with exponent $\xS$ (a potential flow with
an exponent $\xP$).
As noted in Section~\ref{sect:3D}, the critical value of the scaling
exponent equals unity, $\xc=1$, in both cases.
Therefore, the critical scaling exponent for the dynamo effect 
does not depend on whether or not the solenoidal
and potential components scale differently, and if at least
one of the two exponents is greater than unity, a dynamo is obtained.

\begin{figure}[h]
\centering
\includegraphics[width=\columnwidth]{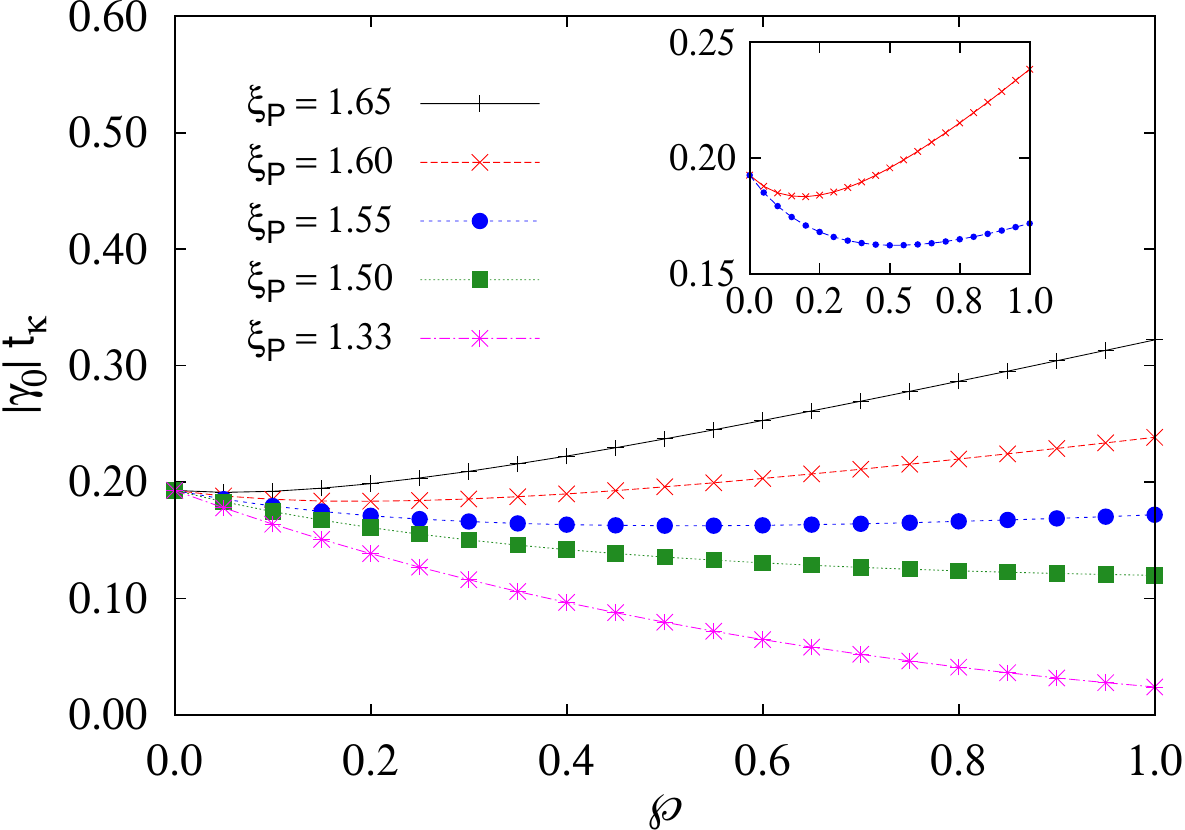}
\caption{Magnetic growth rate multiplied by the diffusive time as a function
of the degree of compressibility of the velocity field for $\xS=4/3$
and different values of $\xP$. The $\xP=1.55$
and $\xP=1.60$ curves are plotted again in the inset, so that
their non-monotonic behaviour is more easily appreciated.}
\label{fig:two-exponents}
\end{figure}

The magnetic growth rate, by contrast, is expected to vary with the spatial regularity
of the potential component. 
Since for a generic $\wp$ the solenoidal and potential components
of the velocity field now scale differently, two scales, 
$r_\kappa^{(\rm S)}\equiv (\kappa/D_{\rm S})^{1/\xi_{\rm S}}$ and 
$r_\kappa^{(\rm P)}\equiv (\kappa/D_{\rm P})^{1/\xi_{\rm P}}$,
can be formed by balancing the diffusion term  
in \eqref{eq:MHD} with
either the solenoidal or the potential contribution to the advection term.
In the calculation of the magnetic growth rate, we set $r_\kappa^{(\rm S)}=
r_\kappa^{(\rm P)}$. The relative weight of the solenoidal and the
potential components is thus determined solely by $\wp$. Furthermore, such 
choice
allows us to define a single diffusive time scale $t_\kappa$, which is used
to rescale the growth rate.
We do not know any analytical method
allowing us to exactly calculate the growth rate as a function of $\xi$ and $\wp$. 
Hence we use the variation--iteration method to calculate it numerically.
In figure~\ref{fig:two-exponents} we plot the non-dimensionalized growth rate
as a function of $\wp$ for $\xS=4/3$ and for several different values of $\xP$. 
According to the discussion at the end of Section~\ref{sec:solran},
a turbulent velocity field with spectrum scaling as in
\cite{F10,F13,Wangetal13,Wangetal18}, 
i.e., $k^{-5/3}$ for the solenoidal component and $k^{-2}$ for the
potential one, corresponds to a 
white-in-time flow with $\xS=4/3$ and $\xP=3/2$ (green square symbols in figure~\ref{fig:two-exponents}). 
The values 
$\xS=\xP=4/3$ correspond to the case where both the solenoidal and the 
potential components of the white-in-time flow
scale with Kolmogorov exponent (red $\ast$ symbols in figure~\ref{fig:two-exponents}). 
In both of these cases the growth rate 
monotonically decreases as a function of $\wp$. 
But the growth rate for the case $\xS=4/3$ and $\xP=3/2$
is higher for all values of $\wp$. 
Remarkably, for greater values of $\xP$, we find
that $\vert\gamma_0\vert$
{\it varies non-monotonically or even increases with} $\wp$ (inset of figure~\ref{fig:two-exponents}).
Hence compressibility can increase the magnetic growth if the potential
component of the flow is sufficiently regular. 
This is a new qualitative result.

\subsection{Critical magnetic Reynolds number} \label{sect:rm_crit}

\begin{figure}
\includegraphics[width=\columnwidth]{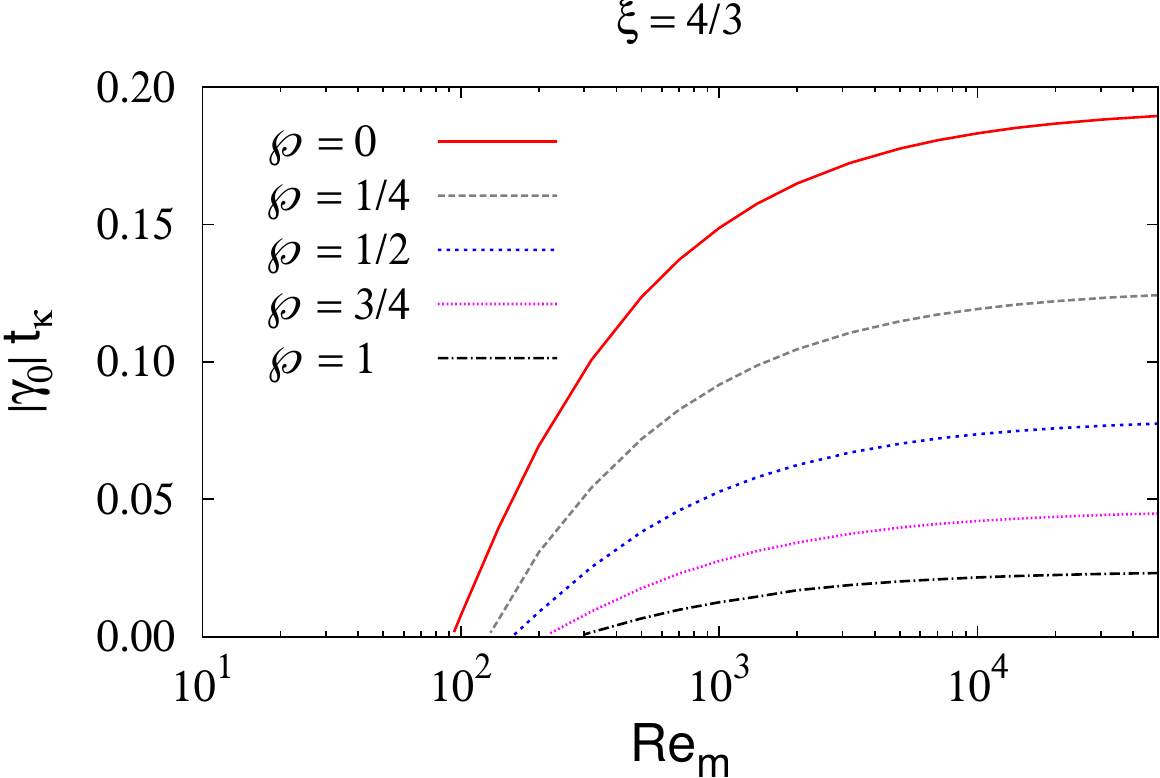}
\\[3ex]
\includegraphics[width=\columnwidth]{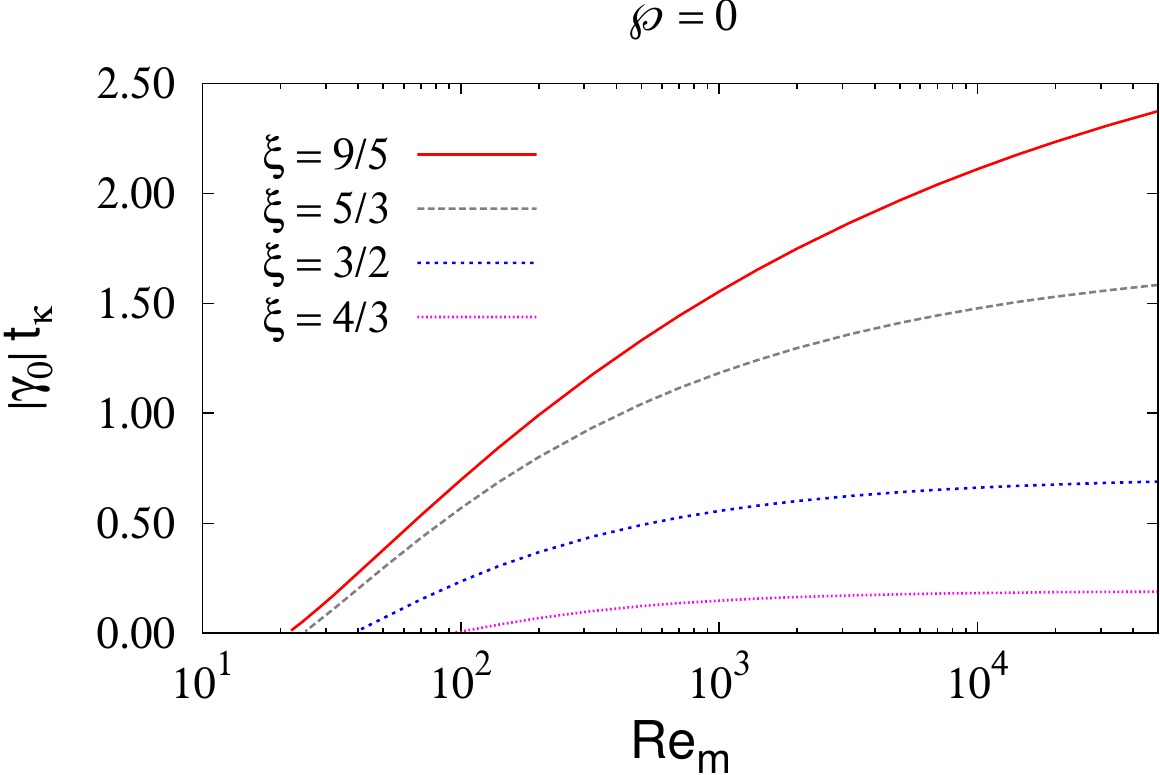}
%\includegraphics[width=0.5\columnwidth]{plot-rm-wp.pdf}%
%\hfill\includegraphics[width=0.5\columnwidth]{plot-rm-xi.pdf}
\caption{Magnetic growth rate multiplied by the diffusive time 
as a function of the magnetic Reynolds number for 
%(left)
(upper panel)
$\xi=4/3$ and from top to bottom $\wp=0,1/4,1/2,3/4,1$, and
%(right)
(lower panel)
$\wp=0$ and from top to bottom $\xi=9/5,5/3,3/2,4/3$.}
\label{fig:Rm_crit_one}
\end{figure}

The magnetic Reynolds number $\mathit{Re}_{\rm m}$ measures
the relative intensity of the stretching of magnetic lines by the flow and
Ohmic dissipation. Therefore, for the dynamo effect to take place,
$\mathit{Re}_{\rm m}$ must be sufficiently high and exceed a critical value
$\mathit{Re}_{\rm m}^{\rm crit}$.

We consider first the case in which both the solenoidal and the potential
components of the velocity field display the same scaling exponent
($\xi_{\rm S}=\xi_{\rm P}=\xi$).
In the original version of the Kazantsev model,
$\mathit{Re}_{\mathrm{m}}$ is infinite, since the structure
functions of the velocity field are assumed to increase indefinitely with the scale separation.
It is possible, however, to modify the form of $\SL(r)$ and $\SN(r)$
in such a way as to introduce a finite correlation length $L$
beyond which the structure functions saturate to
a constant value \cite{RS81,NRS83,AS86,MR87,RSS89,SBK02}.
This allows the definition of the magnetic Reynolds number
as $\mathit{Re_{\rm m}}\equiv L/r_\kappa=L(D/\kappa)^{1/\xi}$.
Here we follow \cite[p.~55]{MY75} (see also \cite{FGV01,V02,AH07}) and introduce $L$ 
in the model by modifying the velocity spectrum as follows:
\begin{equation}
E(k)\propto \dfrac{k^4}{(k^2+L^{-2})^{(5+\xi)/2}}
\end{equation}
(the reader is referred to appendix~\ref{app:d} for more details).
The specific form of $E(k)$ is immaterial and only affects numerical details. 
What is important is
that the power-law behaviour $E(k)\sim k^{-1-\xi}$ is recovered
for $k\gg L^{-1}$, while $E(k)$ is regular for $k\ll L^{-1}$.
The structure functions for finite $\mathit{Re}_{\mathrm{m}}$
can then be calculated from $E(k)$ (see appendix~\ref{app:d}) and can be inserted into 
\eqref{eq:mr} and~\eqref{eq:potential} to obtain the effective potential $\Ueff(r)$,
and hence the magnetic growth rate via the variation--iteration method.

Within the Schr\"odinger formulation of the Kazantsev model,
the existence of a critical magnetic Reynolds number can be interpreted thus: 
a finite value of $L$ generates a potential barrier at 
spatial scales of the order of $L$ and greater than it \cite{NRS83,SBK02,V02}.
As $L$ (and hence $\mathit{Re}_{\rm m}$) is decreased, the potential
well raises, until negative-energy states cease to exist and the dynamo
disappears.

Figure~\ref{fig:Rm_crit_one} shows the rescaled magnetic growth rate as a function of
$\mathit{Re}_{\rm m}$ (a) for fixed $\xi=4/3$ and different values
of $\wp$ and (b) for fixed $\wp=0$ and different values of $\xi$.
We find yet another instance in which 
the degree of compressibility and the spatial regularity of the flow
have an opposite effect on the dynamo effect:
compressibility increases $\mathit{Re}_{\rm m}^{\rm crit}$,
whereas the smoothness of the flow makes it decrease.

It is thus interesting to examine the interplay between these
two effects by considering, as in Sect.~\ref{sect:two-exponents}, a velocity spectrum
such that the solenoidal component of the velocity field
displays the Kolmogorov scaling, while
the potential one the Burgers scaling:
\begin{equation}
\label{eq:spectrum-two-exponents}
\begin{split}
E(k)&=(1-\wp)\frac{A_{\rm S}}{L^{\xi_{\rm S}}}\,
\dfrac{k^4}{(k^2+L^{-2})^{(5+\xi_{\rm S})/2}}
\\
&\quad+\wp \frac{A_{\rm P}}{L^{\xi_{\rm P}}}
\dfrac{k^4}{(k^2+L^{-2})^{(5+\xi_{\rm P})/2}}.
\end{split}
\end{equation}
Here $A_{\rm S}$ and $A_{\rm P}$ are positive coefficients.
The structure functions corresponding to the above spectrum
can be found in appendix~\ref{app:d}.

\begin{figure}
\includegraphics[width=\columnwidth]{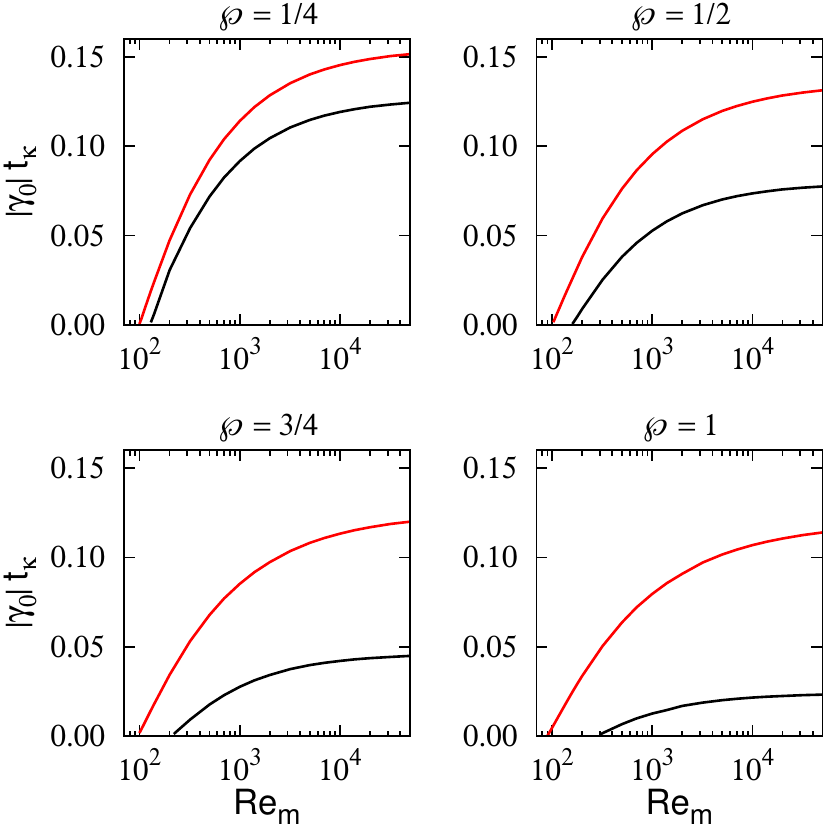}
\caption{
Magnetic growth rate multiplied by the diffusive time 
as a function of the magnetic Reynolds number for different values of $\wp$.
The solid black curves correspond to $\xi_{\rm S}=\xi_{\rm P}=4/3$;
the dashed red ones to $\xi_{\rm S}=4/3$ and $\xi_{\rm P}=3/2$.}
\label{fig:Rm_crit_two_exponents}
\end{figure}

Once more we calculate the magnetic growth rate numerically
by using the variation--iteration method. 
The scales $r_\kappa^{(\rm S)}$ and $r_\kappa^{(\rm P)}$ are defined
as in Sect.~\ref{sect:two-exponents} with $D_{\rm S}$ and $D_{\rm P}$
given in \eqref{eq:ds} and \eqref{eq:dp}.
We again set the ratio $r_\kappa^{(\rm S)}/r_\kappa^{(\rm P)}$ to unity;
the values of $Re_{\rm m}$ and $t_\kappa$ are thus
unaffected by whether $r_\kappa^{(\rm S)}$ or
$r_\kappa^{(\rm P)}$ is used in the definitions.
The magnetic growth rate as a function of $\mathit{Re}_{\rm m}$ is shown 
in figure~\ref{fig:Rm_crit_two_exponents} for the case
$\xi_{\rm S}=4/3$, $\xi_{\rm P}=3/2$ and for
representative values of the degree of compressibility. 
Compared to the case $\xi_{\rm S}=\xi_{\rm P}=4/3$,
the critical
magnetic Reynolds number is lower when the potential
component of the flow displays the Burgers scaling. In this latter case,
$\mathit{Re}_{\rm m}^{\rm crit}$ depends weakly on $\wp$, which suggests
that, for $\xi_{\rm P}=3/2$, the higher spatial regularity compensates
the effect of the increase in compressibility.
For higher values of $\xi_{\rm P}$, the effect of the higher
regularity of the potential component is expected to prevail and to make
$\mathit{Re}_{\rm m}^{\rm crit}$ decrease as a function of $\wp$.

\section{Kinematic dynamo in $\lowercase{d}$ dimensions} \label{sect:general-d}

In this section, we examine the dependence of the 
Kazantsev model (with single scaling exponent) on the space dimension.
In \eqref{eq:velocity}, the dimension $d$
may indeed be regarded as a parameter that can also take values different
from 2 or 3. So far in this paper we have limited ourselves to the
case $d=3$. 
Gruzinov {\em et al.} \cite{GCS96} showed that if the flow is smooth ($\xi=2$) 
and incompressible ($\wp=0$), there is dynamo effect only when the
spatial dimension is in the interval $2.103\leqslant d\leqslant 8.765$.
Arponen and Horvai \cite{AH07} extended this result to a rough flow ($0<\xi<2$)
and showed that the curve $\xc$ vs $d$ is convex and has its minimum 
in $d=3$. Therefore, the range of spatial dimensions over which the 
dynamo effect can take place shrinks as the flow becomes rougher or,
in other words, a higher degree of spatial regularity of the flow
is necessary if $d\neq3$.

\begin{figure}
\centering
\includegraphics[width=\columnwidth]{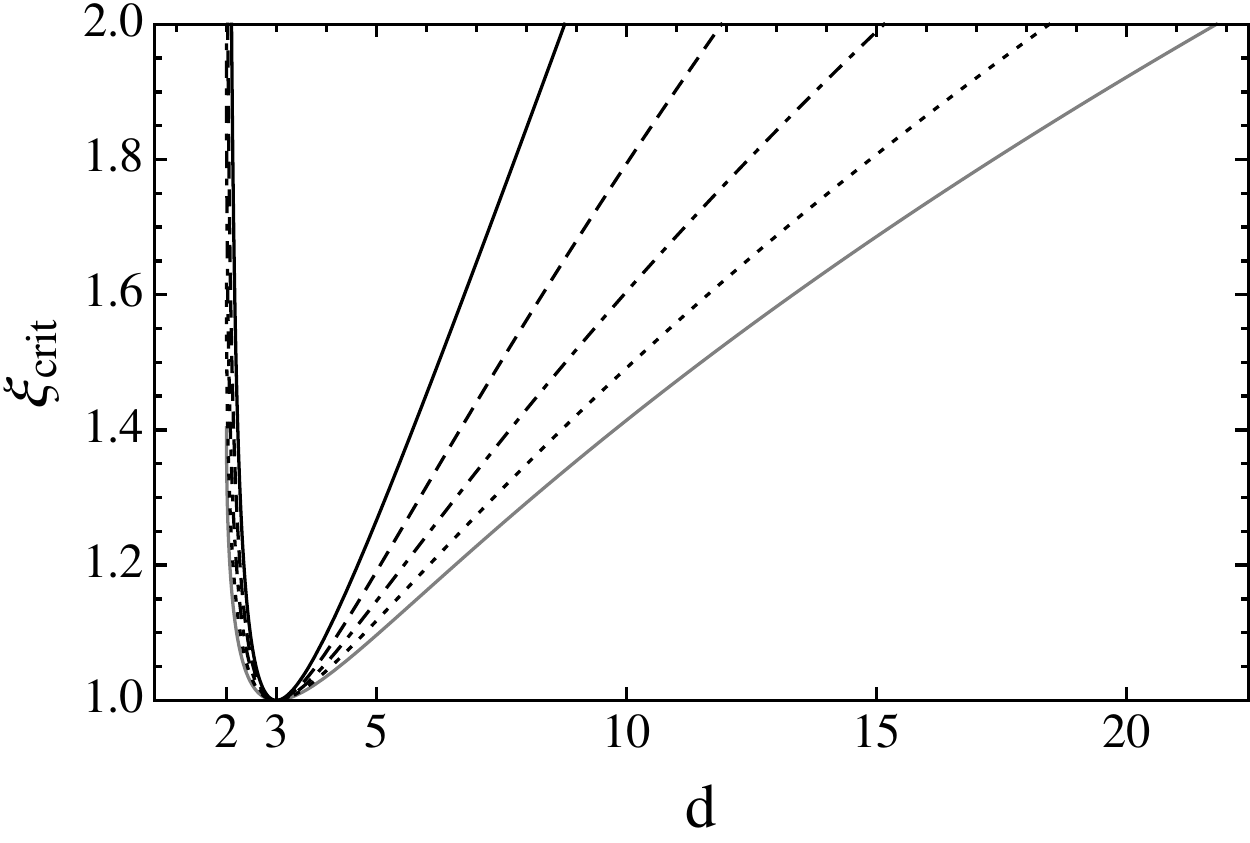}
\caption{The critical scaling exponent for the dynamo effect
vs the spatial dimension of the flow for different values 
of the degree of compressibility. From left to right: 
$\wp=0$ (solid, black line), $\wp=1/4$ (dashed line), 
$\wp=1/2$ (dot-dashed line), $\wp=3/4$ (dotted line), $\wp=1$ (solid, gray line).}
\label{fig:critical-xi}
\end{figure}

To study the effect of compressibility on $\xc$ for a general
dimension $d$, we once again examine the large-$r$ form of $\Ueff(r)$.
By repeating the argument used in Sections~\ref{sect:schroedinger}
and~\ref{sect:3D}, we obtain that $\xc$
is the root of the following equation,
\begin{equation}
\label{eq:below}
\dfrac{a_2}{4[D(d-1)(\wp\xi+1)]^2}=-\dfrac{1}{4},
\end{equation}
such that $0\leqslant \xc\leqslant 2$ (the coefficient 
$a_2$ is defined in \eqref{eq:potential-d-wp} and~\eqref{eq:a2}).
Figure~\ref{fig:critical-xi} 
shows $\xc$ as a function of $d$ for different values of $\wp$.
Clearly, the three-dimensional case is peculiar inasmuch as
it is the only one for which $\xc$ does not depend upon $\wp$,
and the critical exponent is the lowest for $d=3$.
For all other values of $d$, increasing $\wp$ broadens the range
of spatial dimensions over which the magnetic correlation grows, and
lowers the critical exponent for the appearance of the dynamo effect.
No exponential growth of the magnetic field is found
for $d=2$ irrespective of the value of $\wp$.

\section{Conclusions}

In this paper we have used a model that is a generalization of the
Kazantsev model to flows that have both solenoidal and potential
components.
We find, in agreement with earlier results, that the critical value of the exponent above which a
dynamo is seen is unity in all cases. 
If both the solenoidal and the potential parts have the same
scaling exponent, then, as the compressibility of the flow increases, the
growth rate of the dynamo decreases but does not go to zero. 
In other words, even when the flow has only a
potential component, the dynamo is not turned off but merely slowed
down. This qualitative result was already known from the 
approximate analytical work of Rogachevskii and Kleeorin
\cite{RK97}; we have now provided
numerical results that support the same conclusion.
We also show that the flow compressibility 
increases the critical magnetic Reynolds number for the dynamo effect, 
whereas spatial regularity lowers it.
More importantly, we have considered the more realistic case where the
solenoidal and the potential parts have different scaling exponents. 
If we consider the case where these scaling exponents for the solenoidal and
potential parts correspond to typical
Kolmogorov and Burgers values, then we again reach the same 
qualitative conclusion --- an increase in compressibility slows down the
growth rate of the dynamo but does not turn it off. 
The slow down due to compressibility is, however, weaker than 
in the case in which both the solenoidal and the potential components
of the flow display a Kolmogorov scaling.
Intriguingly, we
find that there exist cases, if the potential part is significantly smoother
than the solenoidal part, where an increase in compressibility can actually
{\it increase} the dynamo growth rate. 
Unfortunately, this result remains a curiosity because we are not
aware of any realistic flow that corresponds to such a case. 
If the potential component of the flow displays the Burgers scaling,
we also show that the critical Reynolds number is lower than when both the solenoidal and
the potential components display the Kolmogorov scaling.
Finally, by considering the same model in general $d$ dimensions, we 
realize that the dimension $d=3$ is special; for all other values of
$d$ the critical value of the exponent depends on the degree of
compressibility of the flow, in particular, increasing the
compressibility lowers the critical value of the exponent.

The Kazantsev model is a rare example of a flow allowing
an analytical study of the turbulent dynamo effect.
For this reason, it has attracted
a lot of attention in the literature, and several extensions
of the model have been proposed to include certain
properties of a turbulent velocity field that were not present in 
the original version.
One should of course be careful in translating lessons learned from
Kazantsev-type models to real astrophysical flows
or to direct numerical simulations.
These models are indeed limited by their white-in-time
nature and their Gaussian statistics. Direct quantitative comparison
between such models and simulations may therefore not 
be straightforward. 
Our results are furthermore limited by the fact that we
consider a vanishing magnetic Prandtl number.
We also note that here we have defined the dynamo effect in terms of the
growth or decay of the volume average of the magnetic energy.
In the usual stretch-twist-fold picture of Zeldovich, it is the
magnetic flux that is considered.  It has been shown in earlier works,
see e.g. \cite{CFKV99,MB12,SP18} , that the growth rate of different moments of
the magnetic field can be different.
Therefore, care should be taken in extending our results to
moments of the magnetic field of an order different from the second.
Nevertheless, the large literature on Kazantsev-type models
shows that the qualitative insight gained from the study of such 
models is robust.

\begin{acknowledgments}
The authors are grateful to J\'er\'emie Bec, Axel Brandenburg, 
Antonio Celani, Igor Rogachevskii and Jennifer Schober
for useful suggestions.
M.M.A. was partially supported by CMUP (UID/MAT/00144/2013), funded by FCT (Portugal)
with national (MEC) and European structural funds (FEDER), under the partnership agreement PT2020;
and also by Project STRIDE - NORTE-01-0145-FEDER-000033, funded by ERDF NORTE 2020.
D.M. is supported by grants
from the Swedish Research Council (grant no. 638-2013-9243 and 2016-05225).
\end{acknowledgments}

\appendix

\section{Potential $U$ for general \lowercase{$d$} and $\wp$} \label{app:b}

For general $d$ and $\wp$, the coefficients in the
expression of the potential $U(r)$, see \eqref{eq:potential-d-wp},
are:
\begin{align}
a_0&=4(d+1)(d-1)\kappa^2,
\\
\begin{split}
a_1&=4D[2(\wp\xi^3+\wp\xi^2-\xi^2+2\wp\xi-3\xi+8)
\\
&\quad\ +(\wp\xi^3+2\wp\xi^2-2\xi^2+7\wp\xi-8\xi+20)(d-3)
\\
&\quad\ +2(\wp\xi-\xi+4)(d-3)^2+(d-3)^3],
\end{split}&
\\
\begin{split}
a_2&=D^2[4(\wp^2\xi^4+2\wp^2\xi^3-2\wp\xi^3-4\wp^2\xi^2-4\wp\xi^2-3\xi^2
\\
&\quad\ +4\wp\xi-6\xi+8)+4(\wp^2\xi^4+3\wp^2\xi^3-3\wp\xi^3
\\
&\quad\ -5\wp^2\xi^2-8\wp\xi^2-4\xi^2+9\wp\xi-11\xi+14)(d-3)
\\
&\quad\ +(\wp^2\xi^4+4\wp^2\xi^3-4\wp\xi^3-5\wp^2\xi^2-26\wp\xi^2
\\
&\quad\ -4\xi^2+22\wp\xi-24\xi+36)(d-3)^2\\
&\quad\ -2(3\wp\xi^2-2\wp\xi+2\xi-5)(d-3)^3+(d-3)^4].
\label{eq:a2}
\end{split}&
\end{align}

\section{Variation--iteration method}\label{app:a}

The variation--iteration technique proposed by Morse and Feshbach \cite{MF53}
represents a useful tool for calculating $\psi_{\gamma_0}(r)$
and $\vert\gamma_0\vert$.
It was applied earlier to the study of the
kinematic dynamo effect in the incompressible Kazantsev model
\cite{V02} and is summarized below.

We first note that the eigenfunction $\psi_\gamma(r)$ satisfies the equation:
\begin{equation}
\label{eq:stationary_schroedinger}
\frac{\mathrm{d}^2\psi_\gamma}{\mathrm{d}r^2}+m(r)[\gamma-U(r)]\psi_\gamma=0.
\end{equation}
The interval $[0,\infty)$ is then mapped into the bounded interval
$(0,1]$ by means of the change of coordinate $\rho=(1+\mu)^{-r}$
with $\mu>0$, which transforms \eqref{eq:stationary_schroedinger}
as follows:
\begin{equation}
\mathscr{L}\psi_\gamma=\lambda\mathscr{M}\psi_\gamma,
\end{equation}
with
\begin{equation}
\mathscr{L}\equiv-[\ln(1+\mu)]^2\dfrac{\mathrm{d}}{\mathrm{d}\rho}\left(\rho\dfrac{\mathrm{d}}{\mathrm{d}\rho}\right)
+\dfrac{m(\rho)}{\rho}[U(\rho)-U_{\mathrm{min}}],
\end{equation}
$\mathscr{M}\equiv m(\rho)/\rho$, and $\lambda\equiv \gamma-U_{\mathrm{min}}$.
Here $U_{\mathrm{min}}$ denotes the minimum of the potential $U(\rho)$.
The operators $\mathscr{L}$ and $\mathscr{M}$ are 
positive-definite and self-adjoint with respect to the usual inner product
on $\mathrm{L}^2([0,1])$. In addition, for $\xi>1$,
the spectrum of eigenvalues $\lambda$ is
bounded from below and extends to infinity.

Let $\phi^{(0)}(\rho)$ be an initial trial function that is not orthogonal to
$\psi_{\gamma_0}(\rho)$ and satisfies the boundary conditions
$\phi^{(0)}(0)=\phi^{(0)}(1)=0$. 
No other conditions are required for $\phi^{(0)}(\rho)$.
The technique for calculating
$\psi_{\gamma_0}(r)$ and $\vert\gamma_0\vert$ consists in iteratively
applying the operator $\mathscr{L}^{-1}\mathscr{M}$, in such a way that 
the $n$th iterate $\phi^{(n)}(\rho)$ is given by
\begin{equation}
\label{eq:iteration}
\phi^{(n)}(\rho)\equiv\mathscr{L}^{-1}\mathscr{M}\phi^{(n-1)}(\rho)=
(\mathscr{L}^{-1}\mathscr{M})^n\phi^{(0)}(\rho)
\end{equation}
and the corresponding eigenvalue estimate is:
\begin{equation}
\lambda_0^{(n)}\equiv
\frac{\int_0^1 \phi^{(n)}(\rho)\mathscr{L}\phi^{(n)}(\rho)\mathrm{d}\rho}%
{\int_0^1 \phi^{(n)}(\rho)\mathscr{M}\phi^{(n)}(\rho)\mathrm{d}\rho}.
\end{equation}
As $n$ is increased, $\phi^{(n)}(\rho)$ tends to $\psi_{\gamma_0}(\rho)$
and $\lambda^{(n)}_0$ to $\gamma_0-U_{\mathrm{min}}$ (from above).
The iteration is stopped once the desired convergence is obtained.

In numerical calculations, the interval $[0,1]$ is partitioned into
$N$ subintervals of size $\Delta$ and, for a generic function $f(\rho)$,
the differential part of the operator $\mathscr{L}$
is discretized as follows:
\begin{equation}
\begin{split}
\left.\dfrac{\mathrm{d}}{\mathrm{d}\rho}\!\left(\rho\dfrac{\mathrm{d}f}{\mathrm{d}\rho}\right)\!\right|_{\rho_k}\!\!
=&
\dfrac{1}{2}\!\left(\!
\dfrac{\rho_{k+1}f'_{k+1}-\rho_k f'_{k}}{\Delta}
+\dfrac{\rho_{k} f'_{k}-\rho_{k-1} f'_{k-1}}{\Delta}\!\right)\!
\\
=&\dfrac{1}{2\Delta}\!\left(
\rho_{k+1}\dfrac{f_{k+1}-f_k}{\Delta}
-\rho_{k}\dfrac{f_{k}-f_{k-1}}{\Delta}\right.
\\
&\qquad\left.+\rho_{k}\dfrac{f_{k+1}-f_{k}}{\Delta}
-\rho_{k-1}\dfrac{f_{k}-f_{k-1}}{\Delta}\right)\!
\\
=&\dfrac{1}{2\Delta^2}[(\rho_{k+1}+\rho_{k})f_{k+1}
+(\rho_{k}+\rho_{k-1})f_{k-1}
\\
&\qquad-(\rho_{k+1}+\rho_{k-1}+2\rho_k)f_{k}],
\end{split}
\end{equation}
where $\rho_k=k\Delta$, $f_k=f(\rho_k)$
and $f'_k=f'(\rho_k)$.
The above discretization allows us to preserve
the vanishing boundary conditions on $\phi^{(n)}(\rho)$
at each step of the iteration.
The functions $\phi^{(0)}(\rho)$ and $\phi^{(n)}(\rho)$ 
in \eqref{eq:iteration}
are replaced
with the $N$-dimensional arrays 
$\phi^{(0)}(\rho_k)$ and
$\phi^{(n)}(\rho_k)$,
$k=1,\dots,N$, while the operators $\mathscr{M}$ and $\mathscr{L}$
with the $N\times N$ matrices $M_{ij}=\delta_{ij}m(\rho_i)/\rho_i$
and
\begin{equation}
\begin{split}
\Lambda_{ij}=\delta_{ij}\dfrac{m(\rho_i)}{\rho_i}
[U(\rho_i)-U_{\mathrm{min}}]
+\dfrac{[\ln(1+\mu)]^2}{2\Delta^2}\times
\\
\times
\begin{cases}
(-\rho_{i-1}-\rho_i) & j=i-1
\\
(\rho_{i-1}+2\rho_i+\rho_{i+1}) & j=i
\\
(-\rho_{i}-\rho_{i+1}) & j=i+1
\\
0 & \text{otherwise},
\end{cases}
\end{split}
\end{equation}
where $i,j=1,\dots,N$ (no implicit sum on $i$).
Likewise, the $n$th approximation to the lowest eigenvalue is calculated as:
\begin{equation}
\label{eq:eigenvalue}
\lambda_0^{(n)}\approx
\frac{\sum_{i,j=1}^N \phi^{(n)}(\rho_i)\Lambda_{ij}\phi^{(n)}(\rho_j)}%
{\sum_{i,j=1}^N\phi^{(n)}(\rho_i)M_{ij}\phi^{(n)}(\rho_j)}.
\end{equation}
To summarize, one starts from a trial array $\phi^{(0)}(\rho_k)$
satisfying the appropriate boundary conditions,
repeatedly applies the matrix $\Lambda^{-1}M$ and calculates
$\lambda_0^{(n)}$ from \eqref{eq:eigenvalue},
until $\vert(\lambda^{(n)}_0-\lambda^{(n-1)}_0)/\lambda^{(n)}_0\vert <\epsilon$
for a given $\epsilon$. The array $\phi^{(n)}(\rho_k)$ is renormalized
after every iteration.

The above procedure involves the parameters $N$, $\mu$ and $\epsilon$.
$N$ and $\mu$ must be chosen in such a way as to 
accurately resolve $\Ueff(\rho)$, namely the repulsive barrier at $\rho=1$,
the potential well near to the diffusive scale, and the decay of the potential 
to zero
at $\rho=0$. In particular, the choice of the values of $N$ and $\mu$ requires some
care for $\xi$ near to 2, 
since the effective potential decays to zero slowly in that case.
In our calculations
we varied $N$ between $5\times 10^4$ and $10^7$
and $\mu$ between $10^{-5}$ and $5\times 10^{-2}$ according to
the values of $\xi$ and $\wp$.
The threshold for the convergence, $\epsilon$, needs to be 
sufficiently small, especially for
$\xi$ or $\wp$ close to 1, because the convergence to 
the theoretical eigenvalue may be rather slow for these values
of $\xi$ and $\wp$. In our calculations
$\epsilon$ was varied between $10^{-14}$ and $10^{-8}$.

\section{Structure functions for a finite $R\lowercase{e}_{\lowercase{\mathrm{m}}}$} \label{app:d}

As mentioned in Sect.~\ref{sect:introduction},
the longitudinal and normal three-dimensional spectra of an
isotropic random vector field (see \cite[eq.~(12.31)]{MY75} 
for the definitions)
must be nonnegative.
When constructing the spatial correlation of such a field it is therefore
easier to first define the spectra according to the desired model
and then derive the corresponding 
spatial correlations directly from these spectra.

Following \cite{MY75}, we thus introduce a finite $\mathit{Re}_{\rm m}$
in the Kazantsev model by assuming that the velocity spectrum 
has the form given in \eqref{eq:spectrum-two-exponents}.
Such a spectrum 
can be obtained by choosing the longitudinal and normal three-dimensional 
velocity spectra
as follows:
\begin{align}
\begin{split}
\FL(k)=\wp\,\dfrac{A_{\rm P}}{2\pi L^{\xi_{\rm P}}}\,\dfrac{k^2}{(k^2+L^{-2})^{(5+\xi_{\rm P})/2}},
\end{split}
\\
\begin{split}
\FN(k)=(1-\wp)\dfrac{A_{\rm S}}{4\pi L^{\xi_{\rm S}}}\,\dfrac{k^2}{(k^2+L^{-2})^{(5+\xi_{\rm S})/2}}.
\end{split}
\end{align}
The structure functions corresponding to the above spectra 
are \cite[p.~108]{MY75}:
\begin{align}
\SL(r)&=(1-\wp)\SL^{(\rm S)}(r)+\wp\SL^{(\rm P)}(r),
\\
\SN(r)&=(1-\wp)\SN^{(\rm S)}(r)+\wp\SN^{(\rm P)}(r)
\end{align}
with
\begin{align}
\SL^{(\rm S)}(r)&= 
\frac{A_{\rm S}\sqrt{\pi } }{4 \Gamma \big(\frac{5+\xi_{\rm S}}{2}\big)}
\bigg[\Gamma\Big(\frac{\xi_{\rm S}}{2}\Big)
-2^{1-\frac{\xi_{\rm S}}{2}}\!\left(\frac{r}{L}\right)\!^{\frac{\xi_{\rm S}}{2}}K_{\frac{\xi_{\rm S}}{2}}\!\left(\frac{r}{L}\right)\!\bigg],
\\
\SN^{(\rm P)}(r)&= 
\frac{A_{\rm P}\sqrt{\pi } }{4 \Gamma \big(\frac{5+\xi_{\rm P}}{2}\big)}
\bigg[\Gamma\Big(\frac{\xi_{\rm P}}{2}\Big)
-2^{1-\frac{\xi_{\rm P}}{2}}\!\left(\frac{r}{L}\right)\!^{\frac{\xi_{\rm P}}{2}}K_{\frac{\xi_{\rm P}}{2}}\!\left(\frac{r}{L}\right)\!\bigg],
\end{align}
$\SN^{(\rm S)}(r)=\SL^{(\rm S)}(r)+\frac{1}{2}\,r\,{\rm d}\SL^{(\rm S)}/{\rm d}r$
and $\SL^{(\rm P)}(r)=\SN^{(\rm P)}(r)+r\,{\rm d}\SN^{(\rm P)}/{\rm d}r$
(we remind the reader that our definition of the structure
functions differ from that of \cite{MY75} by a factor of~2). 

For $r\ll L$, $\SL(r)$ and $\SN(r)$ reduce to the functions given in 
\eqref{eq:sl-two-exponents} and
\eqref{eq:sn-two-exponents} with
\begin{align}
\label{eq:ds}
D_{\rm S}&=
\frac{A_{\rm S}\Gamma(-1-\xi_{\rm S})}%
{(3+\xi_{\rm S})L^{\xi_{\rm S}}}\, \cos\frac{\pi\xi_{\rm S}}{2}, 
\\
\label{eq:dp}
D_{\rm P}&=\frac{A_{\rm P}\Gamma(-1-\xi_{\rm P})}%
{(3+\xi_{\rm P})L^{\xi_{\rm P}}}\, \cos\frac{\pi\xi_{\rm P}}{2}.
\end{align}
The structure functions for the case in which the solenoidal and potential 
parts of the velocity field have same scaling exponent can be derived
from the above expressions by setting $\xi_{\rm S}=\xi_{\rm P}=\xi$.

\end{document}